\documentclass[pre,twocolumn,superscriptaddress,floatfix]{revtex4-2}
\usepackage[latin1]{inputenc}
\usepackage{graphicx}
\usepackage{amsmath}
\usepackage{amssymb}
\usepackage{color}
\usepackage[dvips]{epsfig}
\usepackage{epstopdf}
\usepackage{times,mathptmx}
\usepackage{epstopdf}
\usepackage{natbib}

\graphicspath{{figs/}}

\DeclareMathOperator{\arsinh}{arsinh}

\begin{document}


\title{Transport of ions in hydrophobic nanotubes}

\author{Olga I. Vinogradova}
\email[Corresponding author: ]{oivinograd@yahoo.com}
\affiliation{Frumkin Institute of Physical Chemistry and Electrochemistry, Russian Academy of Sciences, 31 Leninsky Prospect, 119071 Moscow, Russia}

\author{Elena F. Silkina}

\affiliation{Frumkin Institute of Physical Chemistry and Electrochemistry, Russian Academy of Sciences, 31 Leninsky Prospect, 119071 Moscow, Russia}

\author{Evgeny S. Asmolov}

\affiliation{Frumkin Institute of Physical Chemistry and Electrochemistry, Russian Academy of Sciences, 31 Leninsky Prospect, 119071 Moscow, Russia}

\date{\today}

\begin{abstract}
The theory of electrokinetic ion transport in cylindrical channels of a fixed surface charge density is revisited. Attention is focused on impact of the hydrophobic slippage and  mobility of adsorbed surface charges. We formulate generalised Onsager relations for a cylinder of an arbitrary radius and then derive exact expressions for the mean electro-osmotic mobility and conductivity. To employ these expressions we perform additional electrostatic calculations, with the special focus on the  non-linear electrostatic effects. Our theory provides a simple explanation of a giant enhancement  of the electrokinetic mobility and conductivity of hydrophobic nanotubes by highlighting the role of appropriate electrostatic and hydrodynamic length scales and their ratios. We also propose a novel interpretation of zeta potentials of cylindrical channels.
\end{abstract}

\maketitle

\section{Introduction}

Nanopores are ubiquitous to nature, and nanoporous materials found applications in various technologies. Electrokinetic transport phenomena in nanotubes include an electro-osmotic flow in response to an applied electric field, a conductance, emerging due to a convective ion transfer by this flow in addition to a conventional migration of ions (electrophoresis), and also a streaming current that is generated by pressure gradient if any~\cite{schoch.rb:2008,bocquet.l:2010}. The transport of ions (i.e. streaming and conductivity currents) in these nanoscale systems is of especial interest being important for physiological phenomena, energy harvesting, biosensing and more~\cite{sparreboom.w:2009,daiguji.h:2010,venkatesan.b:2011,xiao.k:2019,hao.z:2020}. Besides, from the measured streaming current the zeta potential of surfaces can be inferred, which is important in colloid and interface science~\cite{hunter.rj:2013,kirby.bj:2004,kamble.s:2022}. Finally, with the advent of nanofluidics there has been considerable interest in an unusually high conductance of electrolyte solutions in confined systems.

Extensive efforts have gone into investigating conductivity in nanotubes and flat-parallel slits experimentally.  \citet{stein.d:2004} found a remarkably high conductivity of dilute solutions in nanoslits that is independent on the salt concentration. The corresponding to a saturation plateau  conductivity augments with the surface charge density~\cite{schoch.rb:2005}. Measurements in nanotubes have shown that the conductivity plateau decreases with the radius~\cite{siria.a:2013}. \citet{balme.s:2015} reported that the height of conductance plateaus is augmented in hydrophobic nanopores. The body of experimental work
investigating the streaming current (including  inferring the zeta potential) is much less than that for the conductivity current, although
there is some literature in this area~\cite{xie.h:2011,salgin.s:2013,xie.y:2014,chen.s:2020}. However, despite a rapid rise of an experimental activity,  ion transport has been given insufficient attention compared with the transport of water. For example, we are unaware of any previous work that has addressed the impact of hydrodynamic (hydrophobic) slippage~\cite{vinogradova.oi:1999,vinogradova.oi:2011,charlaix.e:2005,vinogradova.oi:2003,joly.l:2006,vinogradova.oi:2009} on the streaming current, although the amplification of an electro-osmotic flow in slippery channels was a subject of active experimental research~\cite{muller.vm:1986,churaev.nv2002,audry.mc:2010,dehe.s:2020}.

There is some literature describing  attempts  to provide a satisfactory theory of electric currents emerging in confined electrolytes.
The majority of previous work considered planar geometries (where the exact solution to a non-linear electrostatic problem exists, although can only be expressed in terms of an elliptic integral) and assumed the classical no-slip boundary conditions at the charged walls~\cite{levine.s:1975}.  The conductivity plateau at low salt has been  attributed to the surfaces of a constant charge density~\cite{stein.d:2004}. \citet{bocquet.l:2010} have briefly discussed the expected shift of the conductance plateau due to a hydrodynamic slip, assuming immobile surface charges. Applying some simple scaling arguments these authors propose that this should be $\propto b/\ell _{GC}$, where $b$ is the slip lenth and $\ell _{GC}$ is the Gouy-Chapman length, which is inversely proportional to the surface charge density~\cite{poon.w:2006}. At slippery surfaces adsorbed charges could be mobile, and, therefore, responding to the external electric field as discussed by \citet{maduar.sr:2015} and supported by \emph{ab initio} simulations~\cite{grosjean.b:2019}.
\citet{mouterde.t:2018} derived scaling expressions describing a contribution of a hydrodynamic slip taking into account the mobility of adsorbed ions that later allowed one to interpret the computer simulation data on channel conductivity with physisorbed surface charges~\cite{mangaud.e:2022}.
 Their results, however, apply only  for a situation of channels of a thickness that significantly exceeds the Debye length $\lambda_D$ of the bulk electrolyte.
Recently \citet{vinogradova.oi:2021} proposed an analytical theory of ion conductivity in slippery parallel-plate channels of an arbitrary thickness. These authors derived simple expressions for the mean conductivity of the channel in two regimes, of thick and thin slits, highlighting the role of hydrodynamic slip and mobility of adsorbed surface charges.

Cylinders constitute a more realistic model for artificial nanotubes and real porous materials, but remain much less theoretically understood. The main challenge is related to electrostatic calculations in the cylindrical geometry. Since the exact solution to non-linear electrostatic problem is still unknown, one needs to rely on numerical solutions or approximations. The analytical theory of an electro-osmotic flow in no-slip nanotubes developed by \citet{rice.cl:1965} is restricted to small surface potentials.  Later \citet{levine.s:1975cyl} extended this model for a no-slip cylinder to higher potentials and obtained expressions for streaming and conductivity currents, but only in the integral form. During last few years some approximate analytical expressions for a conductance in cylindrical nanopores have been obtained by postulating either the overall electroneutrality (`Donnan equilibrium')~\cite{biesheuvel.pm:2016,peters.pb:2016},  and also for the `counter-ions only' case (so-called co-ion exclusion approximation)~\cite{uematsu.y:2018}. A slippery nanotube has been considered in \cite{catalano.j:2016}. These authors  proposed an expression relating the conductivity to the integral of an electrostatic potential, assuming the surface charges are immobile. Numerical integration showed that at a given (high) surface charge the conductivity increases with the slip length, but a physical interpretation of this result has not been proposed. \citet{green2021} also studied hydrophobic nanotubes with immobile surface charged and derived some useful expressions for transport coefficients. He showed that the hydrodynamic slip boundary conditions are consistent with the Onsager relations.   \citet{silkina.ef:2019} performed calculations of an electro-osmotic flow in  nanotubes  by applying electro-hydrodynamic slip boundary conditions~\cite{maduar.sr:2015}, i.e. allowing for a mobility of adsorbed charges in response to electric field.
They obtained a rigorous asymptotic result for a situation of strongly overlapping diffuse layers at high surface potential and predicted a parabolic velocity (and potential) profile. This solution, however, is accurate only when $\lambda_D/\ell_{GC} \leq 1$, i.e. when the surface charge is sufficiently small. The authors interpreted the shift (amplification) of the electro-osmotic velocity profile due to slippage and concluded that the mobility of adsorbed charges reduces the effect, but made no attempt to investigate the transport of ions.

In this paper we present the results of a study of ion transport in the hydrophobic nanotubes with the fixed surface charge density. The hydrophobicity implies that the electro-hydrodynamic boundary condition~\cite{maduar.sr:2015} is imposed, i.e. besides hydrophobic slippage we deal with a migration of adsorbed ions in response to an applied electric field.  The theory has the merit of yielding useful analytical results  as well as being very well suited to numerical work. Our work provides new insight into streaming and conductivity currents in the nanotubes, as well as gives an interpretation of their zeta potential.

Our paper is arranged as follows.  In Sec.~\ref{sec:theory}, we define our system and present some general considerations concerning the Onsager relations and transport coefficients. Section~\ref{sec:Onsager} describes the derivation of the Onsager relations for a hydrophobic nanotube. A detailed calculation of coefficients for the mobility matrix is presented in Sec.~\ref{sec:mobility}. The concept of a zeta potential of hydrophobic nanotubes is introduced. The approximate equations for zeta potentials and mean conductivity are derived in Sec.~\ref{sec:approximate}.
Numerical results for zeta-potentials and conductivity are presented in Sec.~\ref{sec:results} and compared with approximate
theoretical expressions.
We conclude in Sec.~\ref{sec:conclusion}. In Appendix~\ref{app:A} we briefly discuss the possible reduction of ion mobilities with salt concentration to argue that this effect can safely be neglected in our system. Appendix~\ref{app:B} contains a derivation of electrostatic relationships required for calculations of streaming and conductivity currents.

\section{General theory}\label{sec:theory}

\begin{figure}[t]
	\includegraphics[width=0.9\columnwidth]{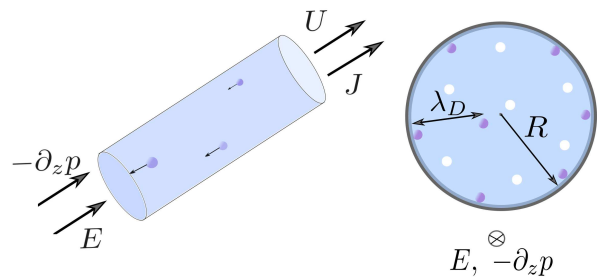}
	\caption{Sketch of the charged nanotube of radius $R$ that is coupled with a
bulk electrolyte reservoir characterized by the Debye length $\lambda_D$. An applied pressure gradient $-\partial_z p$ induces a
hydrodynamic flow and streaming current. An applied electric field $E$ induces an electro-osmotic velocity and a conductivity current. The confined electrolyte flows with the mean velocity $U$ and  has the mean current density $J$.  }
	\label{fig:sketch}
\end{figure}

We consider  an 1:1 aqueous electrolyte solution of  a dynamic viscosity $\eta $
and permittivity $\varepsilon $ in contact with a charged nanotube of radius $R$ as sketched in Fig.~\ref{fig:sketch}. The local number density (per unit volume) of cations and anions are $n_+$ and $n_-$, respectively, and, therefore, the charge densities at each point are $\rho_{\pm} = e n_{\pm}$, where $e$ is the elementary positive charge. Defining $n_{\infty}$ as the number density in the bulk one can write  for the  Boltzmann distribution of density $n_{\pm }(r)=n_{\infty}\exp (\mp \phi (r))$, where $\phi (r)=e\Phi(r)/(k_{B}T)$ is the dimensionless electrostatic
potential, where $k_{B}$ is the Boltzmann
constant, $T$ is a temperature of the system.

The cylindrical system $(z,r)$ of coordinates is defined, such that $z$ coincides with the axis of the nanotube, and $r$ is the radial axis. The surface of nanotube is thus located at $r=R$ and we denote its potential as $\Phi_{s}$ and a charge density as $\sigma $. Without loss of generality, the surface charges are taken as cations ($\sigma$ is positive).

\subsection{Onsager relations}\label{sec:Onsager}

In the general case, the nanotube subject to a pressure gradient $\partial _{z}p$ and an electric field $E$ in the $z$ direction. Thus
the flow inside the nanotube satisfies the Stokes equation, which includes an electrostatic body force $E \rho$
\begin{equation}  \label{eq:Stokes0}
\eta \nabla^2 u=\partial _{z}p+E \rho,
\end{equation}
where $\nabla^2 = \partial_{r}^2 + r^{-1}\partial_{r}$ for the cylindrical geometry, $u(r)$ is the fluid velocity and $\rho$ is the total charge density at each point $r$. It follows from the Poisson equation that
\begin{equation}
  E \rho = - E \dfrac{\varepsilon \nabla^2 \Phi}{4 \pi}.
\end{equation}
Note that CGS (Gaussian) electrostatic units are used throughout our paper. However,
	by expressing all characteristic lengths in terms of the introduced below Bjerrum and Debye lengths we will obtain the results that are independent of any specific system of units.

If the bulk reservoir represents a 1:1 salt solution of concentration $c_{\infty}$,  then
\begin{equation}
  E \rho = - E \dfrac{\varepsilon k_B T \nabla^2 \phi}{4 \pi e} = -   \dfrac{e E \nabla^2 \phi}{4 \pi \ell _{B}},
\end{equation}
where $\ell _{B}=\dfrac{e^{2}}{\varepsilon k_{B}T}$ is  the Bjerrum
length. Note that $\ell _{B}$ of water is equal to about $0.7$ nm for room temperature.

Consequently, Eq.\eqref{eq:Stokes0} can be rewritten as
\begin{equation}  \label{eq:Stokes}
\eta \nabla^2 u=\partial _{z}p+\dfrac{eE}{4\pi \ell _{B}}\nabla^2 \phi
\end{equation}
We assume weak field, so that in steady state $\phi(r)$ is independent on the
fluid flow and satisfies the nonlinear Poisson-Boltzmann
equation
\begin{equation}  \label{eq:NLPB}
\nabla^2 \phi =\dfrac{1}{r} \dfrac{d}{d r} \left( r \dfrac{d \phi}{d r} \right) = \lambda _{D}^{-2}\sinh \phi
\end{equation}
where $\lambda_{D}=\left( 8\pi \ell _{B}n_{\infty}\right) ^{-1/2}$ is the Debye
screening length of an electrolyte solution. Note that by analysing the experimental data it is more convenient to use the concentration $c_{\infty}[\rm{mol/L}]$, which is related to $n_{\infty} [\rm{m^{-3}}] $
as $n_{\infty} = N_A \times 10^3 \times c_{\infty}$, where $N_A$ is Avogadro's number.
We also recall that a useful formula for 1:1 electrolyte is~\cite{israelachvili.jn:2011}
\begin{equation}\label{eq:DLength}
\lambda_D [\rm{nm}] = \frac{0.305 [\rm{nm}]}{\sqrt{c_{\infty}} [\rm{mol/L}]},
\end{equation}
so that upon increasing $c_{\infty}$ from $10^{-7}$ (in pure water, where the ionic strength is due to the dissociating H$^+$ and OH$^-$ ions) to $10^{-1}$ mol/L the screening length is reduced from about 1 $\mu$m down to ca. 1 nm.

To integrate Eq.(\ref{eq:NLPB}) we impose two electrostatic boundary
conditions. The first condition always reflects the symmetry of the channel $%
\phi^{\prime} |_{r=0}=0$, where  $^{\prime}$ denotes $d/d r$. The second condition is applied at the walls
and can be either that of a constant surface potential (conductors) or of a constant surface charge density (insulators). Here we limit ourselves by the constant charge condition, which can be formulated as
\begin{equation}
\phi^{\prime} |_{r=R}=\dfrac{2}{\ell _{GC}},  \label{eq:bc_CC}
\end{equation}
where $\ell _{GC}=\dfrac{e}{2\pi \sigma \ell _{B}}$ is the Gouy-Chapman
length. For high surface charges, say $\sigma \simeq 36 $ mC/m$^2$,  the Gouy-Chapman
length is small, $\ell _{GC} \simeq 1$ nm. However, $\sigma \simeq 0.73$ mC/m$^2$ gives $\ell _{GC} \simeq 50$ nm.

The linearity of Eq.(\ref{eq:Stokes}) implies that its solution represents the decoupled and superimposed  contribution of the pressure-driven and electro-osmotic flows
\begin{equation}
u=-m_h \partial _{z}p+m_{e} E,
\label{eq:vr}
\end{equation}
where $m_h$ and $m_e$ are the functions of $r$ that depend on electrostatic and electro-hydrodynamic boundary conditions at the solid wall. Below we refer these functions to as hydrodynamic and electro-osmotic mobilities. Note that the first term in \eqref{eq:vr} is taken with minus to provide positive $m_h$ when $\partial_{z}p<0$. As we will see below, a sign of $m_e$ depends on the sign of the surface charge (potential).

It is convenient to use the  average values of variables, $U=\overline{u}$, $M_{h}=\overline{m_{h}}$, and $M_{e}=\overline{m_{e}}$.
For any variable $f$ its average value across
the cross-section of the cylindrical nanotube is given by
\begin{equation}
\overline{f}\,=\frac{2}{R^{2}}\int\limits_{0}^{R}\,rfdr.
\end{equation}%

Formally the above considerations are equivalent to the statement that in
the linear response regime,  the transport of water and ions through the nanotube can be then expressed in terms of mobility matrix $\mathcal{M}$:
\begin{equation}
\begin{pmatrix}
U \\
J%
\end{pmatrix}%
=%
\begin{pmatrix}
M_{h} & M_{e} \\
M_{e} & K%
\end{pmatrix}%
\begin{pmatrix}
-\partial _{z}p \\
E%
\end{pmatrix}
= \mathcal{M}
\begin{pmatrix}
-\partial _{z}p \\
E%
\end{pmatrix}
\label{eq:Mo}
\end{equation}%
where $J$ [A/m$^2$] is the mean current density
\begin{equation}\label{eq:generalJs}
  J = \overline{j} + J_{\sigma},
\end{equation}
and $K$ [S/m] is the mean conductivity of the channel.  The first term in \eqref{eq:generalJs} is associated with the transport of the diffuse cations and anions inside the nanotube that generates the local density of current $j = j_{+}+j_{-}$.
The second term reflects the contribution of the surface current $j_{\sigma}$ emerging due to adsorbed mobile ions that are located at the ring of perimeter $2 \pi R$. Therefore, this surface averaged contribution to the mean current density is given by $J_{\sigma} = 2 j_{\sigma}/R$.

As a side note, in experiment one usually measure the conductance $G$ [S]. For a nanotube of length $L$
\begin{equation}
  G = K \dfrac{\pi R^2}{L}
\end{equation}

A $2\times2$ mobility matrix $\mathcal{M}$ is expected to be positive definite and symmetric (with equal off-diagonal coefficients), as assumed in \eqref{eq:Mo}, by analogy with Onsager's relations in (bulk) non-equilibrium thermodynamics. The consequence of the symmetry is so-called electrohydrodynamic coupling
\begin{equation}\label{eq: coupling}
\dfrac{J}{-\partial_{z} p}|_{E=0} = \dfrac{U}{E}|_{-\partial_{z} p=0}.
\end{equation}

The mobility matrix $\mathcal{M}$ fully characterizes electrokinetic phenomena in the nanotubes and, once its elements are known, Eq.\eqref{eq:Mo} can be used to find, without tedious calculations, the liquid flows and currents that are generated by any combination of two applied forces.

However, it is by no means not obvious that by imposing appropriate for hydrophobic surfaces electro-hydrodynamic  boundary conditions, Eq.\eqref{eq:Mo} will necessarily hold. More precisely, the equality of off-diagonal elements $M_e$ is still an assumption that should be proven for a situation when
the fluid velocity at $r=R$ satisfies~\cite{maduar.sr:2015}
\begin{equation}
u|_{r=R}=b\left( - \partial _{r}u|_{r=R}+\dfrac{(1-\mu)eE }{2\pi\eta\ell_{B}\ell _{GC}}\right)  \label{eq:bc_Stokes2}
\end{equation}%
The parameter $\mu$ is the fraction of immobile surface charges
that can vary from 0 for fully mobile charges to 1 in the
case when all adsorbed ions are fixed.
The hydrodynamic slip length $b$ in \eqref{eq:bc_Stokes2} can be of the order of several tens of nanometers~\cite{charlaix.e:2005,vinogradova.oi:2003,joly.l:2006,vinogradova.oi:2009}, but not much more. In our calculations we use $b=10$ nm. However, our discussion will also include a larger value of $b=100$ nm reported in experiment~\cite{vinogradova.oi:2009}.

To prove the consistency of Eq.\eqref{eq:bc_Stokes2} with the  Onsager relations we have to calculate the
coefficients of the mobility matrix. In doing so it is convenient to set one of two possible driving forces to zero.

\subsection{Coefficients of the mobility matrix $\mathcal{M}$}\label{sec:mobility}
\subsubsection{Pressure-driven flow}

If $E = 0$, Eq.~\eqref{eq:Mo} reduces to
\begin{equation}
\begin{pmatrix}
U \\
J%
\end{pmatrix}%
=%
\begin{pmatrix}
M_{h} & M_{e} \\
M_{e} & K%
\end{pmatrix}%
\begin{pmatrix}
-\partial _{z}p \\
0%
\end{pmatrix}%
= -\partial _{z}p \begin{pmatrix}
M_{h} \\
M_{e},%
\end{pmatrix}%
\label{eq:streaming}
\end{equation}
i.e. $U = -M_{h}\partial _{z}p $ (Poiseuille's law), and $J = -M_{e}\partial _{z}p$. In other words, an applied pressure gradient induces not only a hydrodynamic flow, but also an electric current, which is traditionally termed a streaming current.

Local mobilities can be obtained by integrating Eq.~\eqref{eq:Stokes} without an electric body force. The boundary condition \eqref{eq:bc_Stokes2} reduces to a classical hydrodynamic slip boundary condition, and second condition is naturally $\partial _{r}u(r)|_{r=0}=0.$
This gives for a velocity of the pressure-driven flow
\begin{equation}\label{eq:v_pd}
  u = - \dfrac{R^{2}}{4 \eta }\left[1-\dfrac{r^{2}}{R^2}+\dfrac{2b}{R}\right]\partial _{z}p
\end{equation}
The local and mean hydrodynamic mobilities are, therefore,
\begin{equation}\label{eq:vh}
m_{h}=\dfrac{R^{2}}{4 \eta }\left[1-\dfrac{r^{2}}{R^2}+\dfrac{2b}{R}\right], \, M_{h} = \dfrac{R^2}{8\eta} \left( 1 + \dfrac{4b}{R} \right)
\end{equation}

The local densities of current due to diffuse ions are $j_{\pm }=\pm e n_{\pm }u$ with $u$ given by \eqref{eq:v_pd}. Consequently,
\begin{equation}
  \overline{j_{\pm}}   = \pm \dfrac{2e n_{\infty}}{R^2} \int_{0}^{R} r e^{\mp \phi} u dr
\end{equation}
Summing up these functions and using the Poisson-Boltzmann equation \eqref{eq:NLPB}, we conclude that the contribution of the diffuse ions to the density of current can be obtained by taking the integral
\begin{equation}
 \overline{j} =-\dfrac{4en_{\infty }\lambda _{D}^{2}}{%
R^{2}}\int_{0}^{R}ru\nabla ^{2}\phi dr
\end{equation}
Performing (twice) standard integration by parts we then obtain
\begin{equation}\label{eq:j}
 \overline{j} = -\dfrac{e}{4\pi \eta \ell _{B}}\partial _{z}p\left[ \overline{%
\phi }-\phi _{s}-\dfrac{2b}{\ell _{GC}}\right]
\end{equation}

If the adsorbed mobile ions are transferred along the wall with the slip velocity of liquid $u(R) = - \dfrac{b R}{2 \eta}\partial _{z}p$, then they generate the surface current density
\begin{equation}\label{eq:js1}
  j_{\sigma} = \sigma (1-\mu)u(R) = - \dfrac{e (1-\mu) b R }{4 \pi \eta \ell _{GC} \ell _{B}}\partial _{z}p,
\end{equation}
whence
\begin{equation}\label{eq:js}
  J_{\sigma} = - \dfrac{e (1-\mu) b }{2 \pi \eta \ell _{GC} \ell _{B} }\partial _{z}p
\end{equation}

Substituting Eqs.\eqref{eq:j} and \eqref{eq:js} into \eqref{eq:generalJs} we derive for the mean streaming current density
\begin{equation}
 J = - \dfrac{e}{4\pi\eta\ell_{B}}\left( \overline{\phi }-\phi _{s}- \dfrac{2 \mu b}{\ell _{GC}}\right)\partial _{z}p
\end{equation}
The corresponding mean electro-osmotic mobility is
\begin{equation}\label{eq:Me}
 M_{e} = \dfrac{e}{4\pi\eta\ell_{B}}\left( \overline{\phi }-\phi _{s}- \dfrac{2 \mu b}{\ell _{GC}}\right) = - \dfrac{e}{4\pi\eta\ell_{B}} \zeta
\end{equation}

 The new dimensionless parameter $\zeta$ introduced above can be termed a (dimensionless) electro-hydrodynamic or zeta potential
\begin{equation}\label{eq:zeta}
  \zeta = \zeta_0 + \Delta\zeta,
\end{equation}
where
\begin{equation}\label{eq:zeta0}
\zeta_0 = \phi _{s} - \overline{\phi }
\end{equation}
is expected for a hydrophilic nanotube and associated with the (sensitive to amount of added salt) electrostatic contribution,
and
\begin{equation}\label{eq:deltazeta}
\Delta\zeta = \dfrac{2 \mu b}{\ell _{GC}}
\end{equation}
is associated with the slip-driven contribution that does not depend on the salt concentration.

 Note that if we postulate the no-slip boundary condition ($b=0$) and assume that the capillary is infinitely thick (which is equivalent to $\overline{\phi } \to 0$), Eq.~\eqref{eq:zeta} predicts that $\zeta$ must be equal to the surface potential.  However, in the case of slippery nanotubes the situation is more complicated and $\zeta$ does not solely reflect  $\phi_s$, but also depends on the finite average potential in the nanotube and on the charged wall slippage properties.  We shall return to this issue later.

\subsubsection{Electro-osmotic flow}

If $\partial _{z}p =0$, one can reduce Eq.~\eqref{eq:Mo} to
\begin{equation}
\begin{pmatrix}
U \\
J%
\end{pmatrix}%
=%
\begin{pmatrix}
M_{h} & M_{e} \\
M_{e} & K%
\end{pmatrix}%
\begin{pmatrix}
0 \\
E%
\end{pmatrix}%
= E \begin{pmatrix}
M_{e} \\
K%
\end{pmatrix},%
\label{eq:conduction}
\end{equation}
which indicates that an applied electric field $E$ induces both an electro-osmotic flow with the average velocity $U = M_{e}E$ and an electric current of mean density $J = KE$ (Ohm's law). This current is referred below to as a conductivity current.

The electro-osmotic velocity is obtained by integrating  Eq.~\eqref{eq:Stokes} with prescribed boundary conditions (of symmetry and Eq.\eqref{eq:bc_Stokes2})
\begin{equation}\label{eq:ve}
u = \dfrac{e E}{4\pi \eta
	\ell _{B}}\left[\phi -\phi _{s}-\dfrac{2\mu b}{\ell _{GC}}\right],
\end{equation}
whence the local electro-osmotic mobility is
\begin{equation}\label{eq:me}
m_{e}=\dfrac{e}{4\pi \eta
	\ell _{B}}\left[\phi -\phi _{s}-\dfrac{2\mu b}{\ell _{GC}}\right]
\end{equation}
and consequently, $M_e$ is again expressed by Eq.\eqref{eq:Me}. This result provides a proof of Onsager relations \eqref{eq:Mo} if boundary condition \eqref{eq:bc_Stokes2} are applied. We stress, however, that to satisfy the Onsager relations the condition \eqref{eq:js1} for the surface current density should necessary supplement  Eq.\eqref{eq:bc_Stokes2}.

The mean density of the emerging conductivity current is  given by Eq.\eqref{eq:generalJs}, but besides a convective transfer with the fluid velocity that is now described by Eq.\eqref{eq:ve}, ions migrate under an applied field, i.e. they also move relative to a fluid.

The velocity of diffuse ions is thus $\mathcal{U}_{\pm} = u \pm m_{\pm} E$, where $u$ is the velocity of the electro-osmotic flow given by \eqref{eq:ve} and $m_{\pm}$ is the (electrophoretic) mobility of ions. To keep calculations as transparent as possible we
set $m_{\pm} = m = e/(6 \pi \eta \mathcal{R})$, where $\mathcal{R}$ is the hydrodynamic radius of both cations and anions (in calculations below we will use $\mathcal{R} = 0.3$ nm). In such a definition $m$ is treated as equal to the ion mobility at zero ionic strength and we neglect its the possible (weak with our parameters) reduction on increasing $n_{\infty}$ (see Appendix~\ref{app:A}).  If so,
\begin{equation}
  \overline{j_{\pm}} = \pm \dfrac{2e n_{\infty}}{R^2} \int_{0}^{R} e^{\mp\phi} (u \pm m_{\pm} E) r dr
\end{equation}
and consequently
\begin{equation}
\overline{j} =\dfrac{4e n_{\infty }}{R^{2}}%
\int_{0}^{R} r \left(- u \sinh {\phi } + \dfrac{e E}{6 \pi \eta \mathcal{R}} \cosh \phi  \right) dr,
\end{equation}%
which is equivalent to
\begin{equation}
\overline{j} =  K_{\infty } E \left[ - \dfrac{3 \lambda_D^2 \mathcal{R}}{
	\ell _{B} R^{2}}
\int_{0}^{R}  m_e \nabla^2 \phi  r dr +  \overline{\cosh \phi} \right],
\end{equation}%
where
\begin{equation}\label{eq:bulkK}
K_{\infty } = \dfrac{e^2 n_{\infty }}{3 \pi \eta \mathcal{R}},
\end{equation}
is the conductivity of the bulk electrolyte solution.

Substituting expression \eqref{eq:me} for $m_e$ and integrating twice by parts we obtain

\begin{equation}\label{eq:j2}
\overline{j}=K_{\infty } E \left[ \overline{\cosh {\phi }}+\dfrac{3\lambda _{D}^{2}\mathcal{%
R}\overline{(\phi ^{\prime })^{2}}}{2\ell _{B}} + \dfrac{12 \mu b \lambda_D^2 \mathcal{R}}{R \ell _{GC}^{2}\ell _{B}} \right]
\end{equation}

The adsorbed mobile ions generate the density of current
\begin{eqnarray}
  j_{\sigma} &=& \sigma (1-\mu)[u(R)+ mE]\nonumber\\
  &=& \dfrac{(1-\mu )e }{2  \pi \ell _{GC}\ell _{B}} \left( - \dfrac{e E}{2\pi \eta
\ell _{B}} \dfrac{ \mu b}{\ell _{GC}} + \frac{e E}{6 \pi \eta \mathcal{R}}\right),
\end{eqnarray}
where we have substituted $u(R) = - \dfrac{e E \mu b}{2\pi \eta
	\ell _{B}\ell _{GC}}$. Standard manipulations lead to

\begin{equation}\label{eq:jsigma2}
J_{\sigma}  =  K_{\infty } E  \left( \dfrac{4  \lambda_D^2(1-\mu )}{ \ell_{GC}} - \dfrac{12 \mathcal{R}(1-\mu ) \lambda_D^2 \mu b}{
  \ell _{GC}^2 \ell _{B}}   \right)
\end{equation}

Summing up Eqs.\eqref{eq:j2} and \eqref{eq:jsigma2} we find the mean density of the conductivity current $J$, and then by dividing it by $E$ obtain the expression for the mean conductivity
\begin{equation}\label{eq:J2}
K=K_{\infty } \left[ \overline{\cosh {\phi }}+\dfrac{3\lambda _{D}^{2}\mathcal{%
R}\overline{(\phi^{\prime })^{2}}}{2\ell _{B}} + \dfrac{4\ell_{Du}}{ R} \left(  \dfrac{3 \mathcal{R} \mu^2 b}{\ell_{GC} \ell_{B}} + 1 - \mu \right) \right],
\end{equation}
where we use another electrostatic length scale
\begin{equation}  \label{eq:dukhin_length}
\ell_{Du} =  \frac{\lambda_{D}^2}{\ell_{GC}}
\end{equation}
termed the Dukhin length. Note that $\ell_{Du}$ can be larger than any conceivable Debye
length. For example, the values of $\lambda_{D} = 50$ nm and $\ell_{GC} = 2$ nm lead to $\ell_{Du} = 1.25$ $\mu$m, which for a nanotube of $R=10$ nm gives $\ell_{Du}/R = 125$.

It is convenient to divide the mean conductivity of the channel given by \eqref{eq:J2} into a ``no-slip'' conductivity $K_0$ expected for hydrophilic channels, and slip-driven contribution $\Delta K$:

\begin{equation}
K = K_0 + \Delta K,
\label{eq:M22}
\end{equation}
where
\begin{equation}\label{eq:K0}
K_0 = K_{\infty }\displaystyle \left[ \dfrac{3 \lambda _{D}^{2}\mathcal{R}\overline{(\phi^{\prime} )^{2}}}{2\ell _{B}}+ \overline{\cosh {\phi }}\right],
\end{equation}
and
\begin{equation}\label{eq:dK}
  \Delta K = K_{\infty} \frac{4 \ell_{Du}}{R} \left[\frac{3 \mu^2 b}{\ell_{GC}}\frac{\mathcal{R}}{\ell_{B}} + 1-\mu\right].
\end{equation}
Equation \eqref{eq:K0} is equivalent to the formula for $K_0$ of a plate-parallel channel~\cite{vinogradova.oi:2021}. However, now $\overline{(\phi^{\prime} )^{2}}$ and $\overline{\cosh \phi}$ should be calculated for a cylindrical geometry. The form of Eq.\eqref{eq:dK} is identical to that for $\Delta K$ of a plate-parallel channel~\cite{vinogradova.oi:2021}. The only difference is that Eq.\eqref{eq:dK} includes the nanotube radius $R$ instead of the channel thickness.

\section{Approximate solutions}\label{sec:approximate}

In the previous section we have proven the Onsager relations and obtained the general equations for the mean electro-osmotic mobility and the mean conductivity of the nanotube. In order to employ them a detailed  information concerning mean electrostatic parameters, such as $\overline{\phi}$, $\overline{\cosh \phi}$, etc,  is required. The derivation of equations that determine them is given in Appendix~\ref{app:B}.

\subsection{Electro-osmotic mobility and zeta potential}\label{sec:approximate_zeta}

From Eq.~\eqref{eq:Me} it follows that to calculate the electro-osmotic mobility it is enough to find the zeta potential, which incorporates all geometry and electrostatic parameters of the problem.

For a thick cylinder, Eq.\eqref{eq:grahame} for $\phi_s$ of isolated surface remains accurate.
Substituting this equation into \eqref{eq:zeta0} and making the additional assumption that $\overline{\phi} \simeq 0$ one obtains
\begin{equation} \label{eq:Me0_thick}
\zeta_0 \simeq 2 \arsinh(\frac{\lambda_{D}}{\ell_{GC}})
\end{equation}
An approximate $\zeta_0$ given by Eq.\eqref{eq:Me0_thick} may slightly overestimate the exact value since in the cylinder of finite $R$ the average potential does not vanish in the general case (see Appendix~\ref{app:B}).
For weakly charged surface Eq.\eqref{eq:Me0_thick} reduces to
\begin{equation} \label{eq:Me0_thick2}
\zeta_0 \simeq \dfrac{2 \lambda_{D}}{\ell_{GC}}
\end{equation}
All together, $\zeta_0$ of a thick cylinder always depends on salt concentration, and is approximately equal to $\phi_s$.

In the thin-channel regime, substituting \eqref{eq:grahame_thin} and \eqref{eq:averyge_phi_thin} into Eq.\eqref{eq:zeta0} we derive
\begin{equation}\label{eq:Me0_thin}
\zeta_0 \simeq  \dfrac{R}{2 \ell_{GC}}
\end{equation}
Thus, $\zeta_0$ of a thin nanotube is not equal to $\phi_{s}$ given by \eqref{eq:grahame_thin} and depends only on the surface charge and the nanotube radius. Importantly, the zeta potential of a thin hydrophilic cylinder does not depend on the salt concentration, although $\phi_s$ given by \eqref{eq:grahame_thin} reduces on increasing $c_{\infty}$.

The additional slip-driven contribution $\Delta \zeta$ in Eq.\eqref{eq:zeta} does not depend on salt and is of the same form for a
cylinder of any thickness and is given by Eq.\eqref{eq:deltazeta}. It is instructive to calculate the relative enhancement of zeta potential $\zeta/\zeta_0 = 1 + \Delta \zeta/\zeta_0$ due to slippage. Say, straightforward calculations give for a  thick cylinder
\begin{equation}\label{eq:zeta01}
  \dfrac{\zeta}{\zeta_0} \simeq 1 + \dfrac{\mu b}{\ell _{GC} \arsinh(\dfrac{\lambda_{D}}{\ell_{GC}})}
\end{equation}
When the nanotube is weakly charged, Eq.~\eqref{eq:zeta01} reduces to
\begin{equation}\label{eq:zeta02}
  \dfrac{\zeta}{\zeta_0} \simeq  1 + \dfrac{\mu b}{\lambda_D},
\end{equation}
which indicates that an enhancement of zeta potential can become very large at high salt.

For a thin cylinder
\begin{equation}\label{eq:zeta03}
  \dfrac{\zeta}{\zeta_0} \simeq 1 + \dfrac{4 \mu b}{R}
\end{equation}
is defined only by the ratio $\mu b/R$ and does not depend on salt.
To give
an idea on possible zeta potential enhancement, with $\mu b = 100$ nm and $R=10$ nm, we get ca. 40 times amplification!

\subsection{Conductivity}

The substitution of Eqs.\eqref{eq:average_cosh} and \eqref{eq:average_dphi} into Eq.~\eqref{eq:K0} allows one to immediately obtain an expression for electrical conductivity in the thick hydrophilic cylinder
\begin{equation}\label{eq:hydrophilic_thick}
K_0 \simeq  K_{\infty } \left[ 1 + \dfrac{4\lambda_D}{R }\left( \sqrt{1+\dfrac{\ell _{Du}}{\ell _{GC}}}-1\right)\left(1 + \dfrac{3 \mathcal{R}}{ \ell _{B}} \right) \right].
\end{equation}

In the thin-channel regime,  $\overline{\cosh {\phi }}$ and $\overline{(\phi^{\prime} )^{2}}$ are given by Eqs.\eqref{eq:series_dphi_thin} and \eqref{eq:average_cosh_thin}, respectively. Substituting them to \eqref{eq:K0} we derive

\begin{equation}\label{eq:hydrophilic_thin}
K_0 \simeq K_{\infty } \left[\sqrt{ 1 + \left( \dfrac{4\ell_{Du}}{R} \right)^2 } - \dfrac{2 \ell_{Du}}{\ell _{GC}} \left(1 - \dfrac{3 \mathcal{R}}{2 \ell_{B}} \right)\right],
\end{equation}
where the first term is of the leading-order. The significant deviations from the bulk conductivity are expected when $\ell_{Du}/R \geq 1$. In this case $K_0$ does not depend on $c_{\infty}$:
\begin{equation}\label{eq:hydrophilic_thin_plateau}
K_0 \simeq K_{\infty }\dfrac{4\ell_{Du}}{R}
\end{equation}  In other words, the salt dependence of $K_0$ should demonstrate a conductivity plateau in dilute solutions.

In the case of  hydrophobic channels the conductivity can be
obtained from Eq.~\eqref{eq:M22}, i.e. by summing up the corresponding $K_0$ and $\Delta K$ given by \eqref{eq:dK}.
The conductivity amplification due to slippage, $K/K_0 = 1 + \Delta K/K_0$, can then be easily found. For example, if $\ell_{Du}/R\geq 1$ and $\lambda_D/\ell_{GC}$ is small enough, i.e. in the thin channel regime, it follows from Eq.~\eqref{eq:hydrophilic_thin_plateau} that
\begin{equation}\label{eq:hydrophobic_thin}
\dfrac{K}{K_0} \simeq   \frac{3 \mu^2 b}{\ell_{GC}}\frac{\mathcal{R}}{\ell_{B}} + 2 - \mu,
\end{equation}
indicating that the enhancement compared to a hydrophilic nanotube can be very large, a few tens of times, provided $\ell_{GC}$ is small (highly charged surfaces) and adsorbed charges are immobile ($\mu = 1$). Interestingly, $K/K_0 = 2$ when $\mu = 0$, i.e. with fully mobile surface charges the mean conductivity remains enhanced compared to that in hydrophilic nanotube. We recall that in this situation the electro-osmotic flow is the same as in a hydrophilic nanotube.

\section{Results and discussion}\label{sec:results}

In order to assess the validity of the above analysis, we perform
a numerical resolution of Eq.\eqref{eq:NLPB} complemented by the boundary
condition \eqref{eq:bc_CC} following the numerical approach developed
by Bader and Asche. Once $\phi$ and corresponding average functions are found numerically (see Appendix A), the streaming and conductivity currents, zeta potential and mean conductivity can be obtained using the expressions derived in Sec.~\ref{sec:theory} and Sec.~\ref{sec:approximate}. All results in this Section are obtained for nanotubes of $R = 10$ nm.

\subsection{Streaming current}

\begin{figure}[h]
\begin{center}
\includegraphics[width=1\columnwidth]{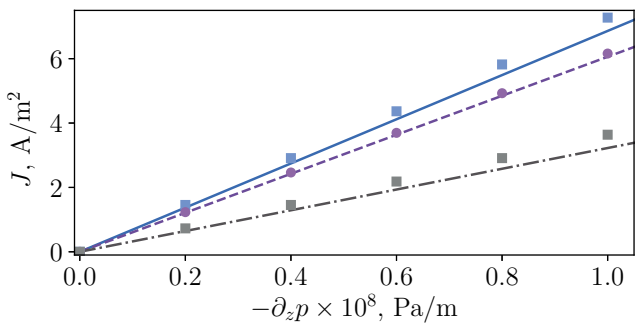}
\end{center}
\caption{Streaming current density $J$ vs. pressure drop $\partial_{z}p$ computed for hydrophobic cylinders of $R = 10$ nm, $b=10$ nm and $\ell_{GC}=5$ nm.  The solid line shows numerical results obtained using $\mu=0.75$ and $c_{\infty} = 10^{-5}$ mol/L. The dashed line is computed for $\mu=0.75$ and $c_{\infty} = 10^{-1}$ mol/L. The dash-dotted line is computed for $c_{\infty} = 10^{-5}$ mol/L and $\mu = 0.25$.  Circles and squares show calculations using $\zeta_0$ given by Eqs.\eqref{eq:Me0_thick} and \eqref{eq:Me0_thin}, respectively. }
\label{fig:streaming_current}
\end{figure}

Figure~\ref{fig:streaming_current} shows the streaming current as a function of applied pressure gradient. The calculations are made for several hydrophobic nanotubes of the same radius $R=10$ nm, surface charge and slip length, but different $\mu$, computed for two concentrations of salt, $c_{\infty} = 10^{-5}$ mol/L and $c_{\infty} = 10^{-1}$ mol/L which corresponds to $\lambda_{D} \simeq 100$ nm and 1 nm, i.e. to the regimes of thin and thick cylinder. The fixed $\ell_{GC}=5$ nm corresponds to a surface charge density $\sigma \simeq 7.3$ mC$/$m$^2$.
The response is always linear, and the slope of the straight lines is invoked to find the electro-osmotic mobility $M_e$ that depends on $c_{\infty}$. The computed data show that for nanotubes of the same $\mu$ a larger $J$ is observed at lower concentration, but a four orders of magnitude increase in $c_{\infty}$ reduce the streaming current only slightly. However, on reducing $\mu$ from 0.75 down to $0.25$ the streaming current is significantly suppressed. Once $M_{e}$ is known, we can determine the value of $\zeta$. Calculations  from Eq.~\eqref{eq:Me} show that the numerical results (from top to bottom) in Fig.~\ref{fig:streaming_current} correspond to $\zeta \simeq 4.7, 4.2$ and 2.2.
Also included in Fig.~\ref{fig:streaming_current} are the streaming currents calculated analytically using $\zeta$ given by \eqref{eq:zeta} with $\zeta_0$ obtained for the thin and thick cylinder limits (Eqs.\eqref{eq:Me0_thick} and \eqref{eq:Me0_thin}). We see that with these parameters Eq.~\eqref{eq:Me0_thick} provides an excellent fit to numerical results, but Eq.~\eqref{eq:Me0_thin} slightly overestimates the results. It must be remembered that this is a first-order calculation only, and given the simplifications made
 we do not expect it to be very accurate.

\begin{figure}[t]
\begin{center}
\includegraphics[width=0.99\columnwidth , trim=0.cm 0. 0.0cm
0.,clip=false]{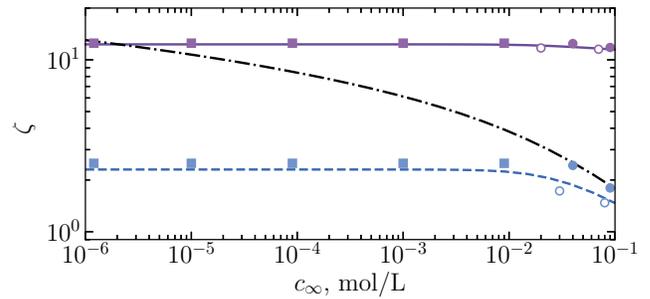}
\end{center}
\caption{Zeta potential as a function of
 $c_{\infty}$ computed for cylindrical channels of $R = 10$ nm, $\ell_{GC} = 5$ nm, $b = 10$ nm using  $\mu = 0.5$ (solid curve) and 0 (dashed curve). The surface potential $\phi_{s}$ is shown by dash-dotted curve.  Filled circles show $\zeta$ calculated using $\zeta_0$ predicted by Eq.\eqref{eq:Me0_thick}, open circles are calculated using $\zeta_0$ given by Eq.\eqref{eq:average_phi_DH2}. Squares are obtained using $\zeta_0$ calculated from Eq.\eqref{eq:Me0_thin}.}
\label{fig:M_cp}
\end{figure}

We now turn to zeta potential of nanotubes.  The main issue we address is how to enhance $\zeta$ by generating a slip velocity at the surface and by tuning the salt concentration. Figure~\ref{fig:M_cp} includes $\zeta$ computed as a function of $c_{\infty}$. Note that varying $c_{\infty}$ from $10^{-6}$ to $10^{-1}$ mol/L is equivalent to the range of $R/\lambda_{D}$ from 0.032 to 10.3. In this calculations we keep $\ell_{GC}=5$
nm fixed, so that $\phi_s$ varies with salt, but the ratio $b/\ell_{GC}=2$ remains constant, i.e. independent on $c_{\infty}$. The computed $\phi_s$ is also included in Fig.~\ref{fig:M_cp}. It can be seen that on increasing $c_{\infty}$ it reduces, but never becomes small. The calculations of $\zeta$ are made using $\mu = 0.5$ and 0. As discussed above, in the latter case of fully mobile surface charges $\zeta = \zeta_0$, although the hydrodynamic slip length is finite. For the thin nanotubes $\zeta$ does not depend on salt and takes its maximal (constant)  value. When nanotubes become thick, $\zeta$ begins to decrease with $c_{\infty}$. It can be seen that when $\mu = 0$ the surface potential is always larger than $\zeta = \zeta_0$, but at $\mu = 0.5$ the zeta potential is significantly amplified by hydrodynamic slippage and in the sufficiently concentrated solutions $\zeta$ dramatically exceeds $\phi_s$. The numerical data are compared with the analytical calculations in which $\zeta_0$ is evaluated using Eqs.\eqref{eq:Me0_thick} (thick nanotubes, $c_{\infty}\simeq 10^{-1}$ mol/L) or \eqref{eq:Me0_thin} (thin nanotubes,   $c_{\infty} \leq 10^{-2}$ mol/L). The fits are quite good, but there is some discrepancy. The analytical equations overestimate the exact numerical results, which is the consequence of an assumption $\overline{\phi} \simeq 0$ (see Sec.~\ref{sec:approximate_zeta}). Similar fits for thick nanotubes made using a next order approximation, Eq.\eqref{eq:average_phi_DH2},  would make an improvement to the fit, but slightly underestimate the numerical results.

\begin{figure}[t]
\begin{center}
\includegraphics[width=0.99\columnwidth , trim=0.cm 0. 0.0cm
0.,clip=false]{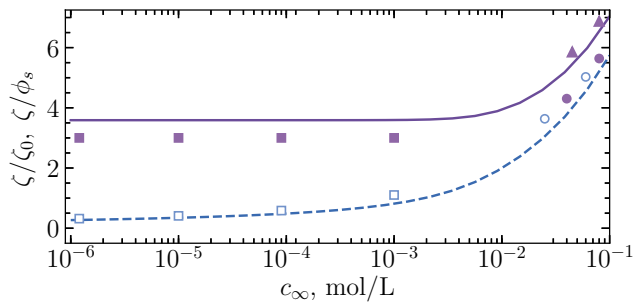}
\end{center}
\caption{$\zeta/\zeta_0$ (solid curve) and $\zeta/\phi_s$ (dashed curve) computed as a function of
 $c_{\infty}$  for cylindrical channels of $R = 10$ nm, $\ell_{GC} = 5$ nm, $b = 10$ nm, and  $\mu = 0.5$. Filled  circles and squares are obtained from \eqref{eq:zeta02} and
  \eqref{eq:zeta03}, respectively. Filled triangles show $\zeta/\zeta_0$ calculated using $\zeta_0$ given by \eqref{eq:average_phi_DH2}. Filled  circles and squares are calculations of $\zeta/\phi_s$ with $\phi_s$ obtained from \eqref{eq:grahame} and \eqref{eq:grahame_thin}.   }
 \label{fig:M_A}
\end{figure}

The amplification of the zeta potential relative to $\zeta_0$  is illustrated in Fig.~\ref{fig:M_A}, where the data are reproduced from Fig.~\ref{fig:M_cp}. With these parameters $\zeta$ of the hydrophobic channel is amplified in ca. 3.5 times in dilute solutions, where $\lambda_D \geq R$. In more concentrated solutions (of small $\lambda_D$) $\zeta/\zeta_0$ begins to rapidly augment with $c_{\infty}$ to larger values. It is instructive to compare these results
with a calculation of $\zeta/\phi_s$, also included in Fig.~\ref{fig:M_A}. This ratio  increases with $c_{\infty}$ strictly monotonically and is always lower than $\zeta/\zeta_0$. When $c_{\infty} \leq 10^{-3}$ mol/L, $\zeta/\phi_s$ is smaller than unity, but then $\zeta$ exceeds $\phi_s$. This is especially well seen in Fig.~\ref{fig:M_cp}. Also included in Fig.~\ref{fig:M_A} are the theoretical results for $\zeta/\zeta_0$ obtained from  Eqs.\eqref{eq:zeta02} and
  \eqref{eq:zeta03}. The theoretical predictions provide a good qualitative agreement with the numerical data, but show smaller amplification of the zeta potential. An obvious explanation for this discrepancy is the neglected average potential of the nanotube. Indeed, Eq.\eqref{eq:average_phi_DH2} that takes into account a non-zero $\overline{\phi}$ for the high concentration branch in Fig.~\ref{fig:M_A} provides a better
match
to the numerical data. If similar fits are made to a $\zeta/\phi_s$ curve, as shown in Fig.~\ref{fig:M_A}, it is found that the agreement of analytical theory that invokes Eqs.\eqref{eq:grahame} and \eqref{eq:grahame_thin} for $\phi_s$ with numerical results is very good.

\subsection{Electrical conductivity}

\begin{figure}[t]
\begin{center}
\includegraphics[width=0.99\columnwidth , trim=0.cm 0. 0.0cm
0.,clip=false]{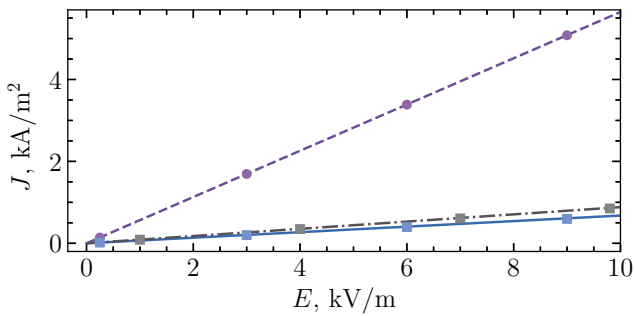}
\end{center}
\caption{Conductivity current density $J$ vs. electric field $E$ computed for the same nanotubes and $c_{\infty}$ as in Fig.~\ref{fig:streaming_current}. Symbols show $J$ calculated using Ohm's law with $K$ given by Eq.\eqref{eq:M22} and $\Delta K$ determined from \eqref{eq:dK}. The circles correspond to $K_0$ calculated from \eqref{eq:hydrophilic_thick}. The squares are calculations using $K_0$ given by \eqref{eq:hydrophilic_thin}.  }
 \label{fig:conduction_current}
\end{figure}

Next, we turn to the mean conductivity of the channel $K$. Figure~\ref{fig:conduction_current} shows a current-voltage response ($J-E$) computed for the same nanotubes and electrolyte concentrations as in Fig.~\ref{fig:streaming_current}. The numerical examples show that $J$ is largest for $c_{\infty} = 10^{-1}$ mol/L. The mean electric current density computed using $c_{\infty} = 10^{-5}$ mol/L is ca. an order of magnitude smaller.
The slope of the $J-E$ straight line is invoked to find the conductivity $K$ at given $c_{\infty}$. Therefore, we conclude that numerical data at $c_{\infty} = 10^{-5}$ mol/L shows conductivity that is an order of magnitude smaller than this computed using $c_{\infty} = 10^{-1}$ mol/L. Meantime, the bulk conductivity is
four order of magnitude smaller as follows from Eq.\eqref{eq:bulkK}. These numerical data are compared with the above theoretical results. The mean conductivity of the nanotube is calculated from Eq.\eqref{eq:M22}. To calculate $K_0$ in this equation in the case of $c_{\infty} = 10^{-1}$ mol/L we use Eq.\eqref{eq:hydrophilic_thick} since $\lambda_D/R$ is small. The calculation of $K_0$ at  $c_{\infty} = 10^{-5}$, where  $\lambda_D/R$ is large, are performed using \eqref{eq:hydrophilic_thin}. For both concentrations to obtain $\Delta K$ we, of course, employ Eq.\eqref{eq:dK}. Once $K$ is known, $J$ can be found employing Ohm's law. Theoretical results are included in Fig.~\ref{fig:conduction_current}. It can be seen that the numerical data sets are very well fitted by analytical equations.

\begin{figure}[t]
\begin{center}
\includegraphics[width=0.99\columnwidth , trim=0.cm 0. 0.0cm
0.,clip=false]{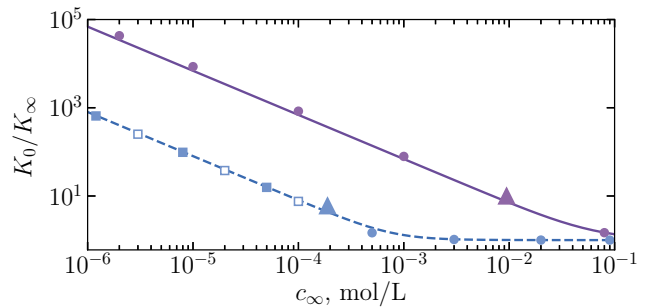}
\end{center}
\caption{$K_0/K_{\infty}$ as a function of
 $c_{\infty}$ computed for nanotubes of $R = 10$ nm, using $\ell_{GC} = 1$ and 50 nm (from top to bottom). Circles show calculations using $K_0$ calculated from Eq.\eqref{eq:hydrophilic_thick}. Filled and open squares are obtained using $K_0$ from Eqs.\eqref{eq:hydrophilic_thin} and \eqref{eq:hydrophilic_thin_plateau}.
 The big triangles mark the points of $\ell_{Du}/{R} = 1$. }
\label{fig:K0_cp}
\end{figure}

To examine the significance of deviations from the bulk conductivity more closely, in Fig.~\ref{fig:K0_cp} we plot $K_0/K_{\infty}$ as a function of $c_{\infty}$. The calculations are made using $\ell_{GC} = 1$ and 50 nm.  The conductivity of hydrophilic nanotubes, $K_0$, can be very large (nearly five orders of magnitude larger than $K_{\infty}$) if it is highly charged ($\ell_{GC} = 1$ nm), but even for $\ell_{GC} = 50$ nm we observe ca. three orders of magnitude enhancement,  provided
that an electrolyte solution is extremely dilute. On increasing $c_{\infty}$ (reducing $\lambda_D$ and $\ell_{Du}$), $K_{0}/K_{\infty}$ first decays linearly in this log-log plot indicating that $K_{0}/K_{\infty} \propto c_{\infty}^{-1}$. However, on increasing $c_{\infty}$ further $K_{0}/K_{\infty}$  begins to reduce weaker (down to 1). We have marked with triangles the points of $\ell_{Du}/{R} = 1$ and one can see that this branch for both curves occurs in the vicinity of these points.
The upper curve that corresponds to a highly charged cylinder is well fitted using $K_0$ given by Eq.\eqref{eq:hydrophilic_thick} indicating that the nanotube effectively behaves as thick even when $\lambda_D/R$ is large. The curve for the weakly charged nanotube is well described using $K_0$ calculated from Eq.\eqref{eq:hydrophilic_thin} when $c_{\infty}$ is below $10^{-4}$ mol/L. The use of \eqref{eq:hydrophilic_thin_plateau} makes practically no difference to the fit, as expected. At concentrations $c_{\infty} \geq 10^{-3}$ mol/L the weakly charged nanotube is effectively thick and the numerical data are well fitted if $K_0$ given by \eqref{eq:hydrophilic_thick} is employed.

\begin{figure}[t]
\begin{center}
\includegraphics[width=0.99\columnwidth , trim=0.cm 0. 0.0cm
0.,clip=false]{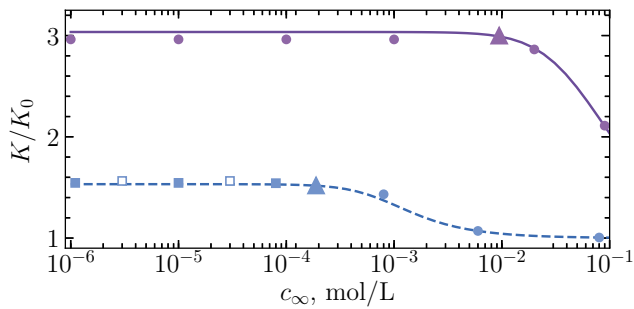}
\end{center}
\caption{The conductivity amplification $K/K_0$ as a function of $c_{\infty}$ obtained for the same nanotubes as in Fig.~\ref{fig:K0_cp}, but with $b = 10$ nm, and $\mu = 0.5$.  Filled symbols are obtained using $K_0$ from Fig.~\ref{fig:K0_cp} and $K$ calculated from \eqref{eq:M22}. Open squares show calculations from Eq.\eqref{eq:hydrophobic_thin}.  The big triangles mark the points of $\ell_{Du}/{R} = 1$. }
\label{fig:K_slip}
\end{figure}

The mean conductivity of the nanotubes can be amplified by the slippage effect. This is illustrated in Fig.~\ref{fig:K_slip}, where the amplification of the mean conductivity is calculated using $\mu = 0.5$ (a half of surface charges is mobile) and a moderate slip length $b = 10$ nm. For these examples we use $\ell_{GC}=1$
and 50 nm, the same as in Fig.~\ref{fig:K0_cp}. It can be seen, that for low concentrations ($\ell_{Du}/{R} \geq 1$) the conductivity amplification due to slip, $K/K_0$, is independent on salt and takes its largest value. For a highly charged nanotube ($\ell_{GC}=1$ nm) the slippage increase the conductivity by a factor of ca. 3, but for a weakly charged cylinder the enhancement is low with these papameters. When $\ell_{Du}/{R}$ becomes smaller than unity, the value of $K/K_0$ decreases with salt. It is well seen that for a curve calculated using $\ell_{GC}=50$ nm $K \simeq K_0$ when $c_{\infty} \geq 10^{-2}$ mol/L. In other words, there is no conductivity amplification due to slippage. However, a highly charged nanotube ($\ell_{GC}=1$ nm) shows the enhancement of the mean conductivity in ca. 2 times even when $c_{\infty} = 10^{-1}$ mol/L. The results of theoretical calculations are also shown in Fig.~\ref{fig:K_slip}. The theoretical data for $K_0$ of  Fig.~\ref{fig:K0_cp} are reproduced and summed up with $\Delta K$ given by \eqref{eq:dK} to obtain $K$. Calculated values of $K/K_0$ fit very well the numerical curves. We have also verified Eq.\eqref{eq:hydrophobic_thin}. As predicted, it represents a sensible approximation for $K/K_0$ if  $\ell_{Du}/R\geq 1$ and $\lambda_D/\ell_{GC} \leq 1$. Indeed, the plateau branch of the numerical curve obtained using $\ell_{GC}=50$ nm is well fitted by Eq.\eqref{eq:hydrophobic_thin}.

\section{Conclusion}\label{sec:conclusion}

We have presented a theory describing the transport of ions in hydrophobic nanotubes with the constant surface charge density. The electro-hydrodynamic boundary condition~\cite{maduar.sr:2015} is imposed at the hydrophobic wall, which assumes that the surface demonstates a hydrodynamic slippage and some portion of the adsorbed surface charges can migrate relative to fluid by  reacting to the applied electric field (but not to the hydrodynamic tangential stress).  Numerical solutions are presented and fully validate our analysis. These results are directly relevant for enhanced streaming and conductivity currents in carbon and boron nitride nanotubes, which are currently the area of very active research, as well as for conventional nanoporous membranes.

The main results of our work can be summarized as follows. We have proven that when the electro-hydrodynamic boundary condition is applied the Onsager relations hold provided the adsorbed surface ions are transferred in a pressure-driven flow with the slip velocity of liquid. Namely, we have derived general expressions for elements of the $2 \times 2 $ mobility matrix, Eq.\eqref{eq:Mo},  and demonstrated that the off-diagonal coefficients are equal if the electro-hydrodynamic boundary condition, Eq.\eqref{eq:bc_Stokes2}, is imposed and provided Eq.\eqref{eq:js1} is valid.
These expressions include some mean electrostatic functions, which we have calculated analytically for two specific regimes (of thin and thick nanotubes). Importantly, these regimes are defined not by just $\lambda_D/R$, but also controlled by the value of $\lambda_D/\ell_{GC}$, i.e. effective charge of the walls. We have then derived simple analytical approximations for the electro-osmotic mobility and mean conductivity in these two regimes. Our results
show that qualitative features of the electrosmotic mobility and conductivity curves for cylinders are the same as for slits, but there is some important quantitative difference due to different expressions for electrostatic functions. We have also given a novel interpretation of the zeta potential of nanotubes.

Our results open strategies to tune the ion transport in nanotubes via a modification of their walls and, vice versa, to probe surface properties by  measuring the streaming or conductivity currents. Our quantitative results can be improved
by performing more accurate calculations of the central (axis) potential. This remains a challenging mathematical problem and a subject of future research. Other fruitful
directions would be to extend the results for so-called charge regulation surfaces and for systems, where both pressure drop and electric field are applied simultaneously. The latter calculations are  now straightforward and can be  done by using the proven Onsager relations.

\begin{acknowledgments}

This work was supported by the Ministry of Science and Higher Education of the Russian Federation.
\end{acknowledgments}

\section*{DATA AVAILABILITY}

The data that support the findings of this study are available
within the article.

\appendix

\section{Mobility of ions in dilute 1:1 electrolyte solutions}\label{app:A}

In general case, the electrophoretic mobility of ions reduces with the concentration of salt. If $\mathcal{R} \ll \lambda_D$, the mobilities of monovalent ions can be approximately described by the H\"uckel formula~\cite{huckel.e:1924}
\begin{equation}\label{eq:huckel}
 m \simeq  \dfrac{e}{6 \pi \eta \mathcal{R}}\left(1 -
  \dfrac{\mathcal{R}}{\lambda_D}\right)
\end{equation}
Using \eqref{eq:DLength} this can be rewritten as
\begin{equation}\label{eq:huckel2}
  m \simeq \dfrac{e}{6 \pi \eta \mathcal{R}}\left(1 -
  \dfrac{\mathcal{R}\sqrt{c_{\infty}}}{0.305}\right)
\end{equation}
This expression shows that the mobility at infinite dilution represents its upper possible limit. The deviations from this ideal mobilty in dilute solutions do not depend on the ion radius and scale with $c_{\infty}^{1/2}$. Thus, Eq.\eqref{eq:huckel2} is functionally equivalent to the Kohlrausch empirical model~\cite{kohlrausch.f:1900}.

It follows from \eqref{eq:huckel2} that at $c_{\infty} = 10^{-2}$ mol/L the decrease in the mobilities of ions of $\mathcal{R} = 0.3$ nm compared to those at an infinite dilution is less than 10\%. Concequently for solutions of $c_{\infty} \leq 10^{-2}$ mol/L (where $\mathcal{R}/\lambda_D$ is small), this effect can safely be neglected. At larger concentrations the analysis based on an upper limit of ion mobilities becomes quite  approximate. Nevertheless, it provides us with some guidance.
 As a side note, at our largest concentration $c_{\infty} = 10^{-1}$ mol/L the use of Eq.\eqref{eq:huckel} would also represent too rough an approximation since $\mathcal{R}/\lambda_D \simeq 0.3$.

\section{Derivation of electrostatic equations}\label{app:B}

\begin{figure}[t]
\begin{center}
\includegraphics[width=0.99\columnwidth , trim=0.cm 0. 0.0cm 0.,clip=false]{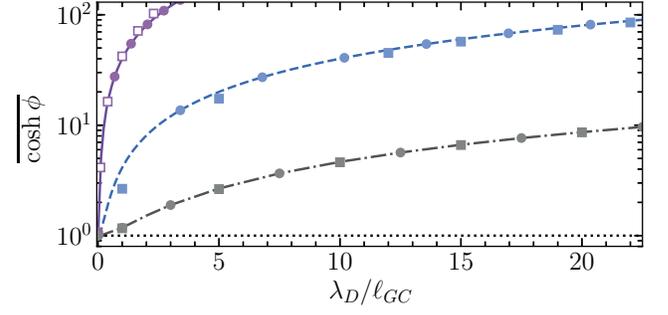}
\end{center}
\caption{$\overline{\cosh\phi}$ as a function of
 $\lambda_D/\ell_{GC}$ computed for cylindrical channels of $R=10$ nm with fixed $\lambda_{D} = 100$ nm (solid curve), $\lambda_{D} = 10$ (dashed curve), and $\lambda_{D} = 1$ (dash-dotted curve). The dotted line indicates $\overline{\cosh\phi}=1$. Circles show predictions of Eq.\eqref{eq:cosh}. Squares are obtained using Eqs.\eqref{eq:average_cosh} and \eqref{eq:average_cosh_thin}.}
\label{fig:cosh}
\end{figure}

Expressions for the mean osmotic pressure $\overline{\cosh {\phi }}$ and for the mean square derivative of the electrostatic potential $\overline{(\phi^{\prime} )^{2}}$, which is the measure of the electrostatic field energy (per unit area), can be derived similar to \cite{vinogradova.oi:2021}.  First integration of Eq.\eqref{eq:NLPB} from the axis ($r=0$) to an arbitrary $r$ gives
\begin{equation}\label{eq:first_int}
\lambda _{D}^{2}\left( \dfrac{d\phi }{dr}\right) ^{2}= 2\cosh \phi -\frac{4}{%
r^{2}}\int_{0}^{r}r\cosh \phi dr
\end{equation}
Applying boundary condition \eqref{eq:bc_CC} then yields
\begin{equation}\label{eq:cosh}
\overline{\cosh\phi} = \cosh\phi_s  - \frac{2 \ell_{Du}}{\ell_{GC}}.
\end{equation}
The expression \eqref{eq:cosh} is exact and valid for any channel thickness and surface charge/potential. In Fig.~\ref{fig:cosh} we present numerical solutions to validate Eq.\eqref{eq:cosh} and illustrate the variation of $\overline{\cosh\phi}$ in response to $\lambda_{D}/\ell_{GC}$ and $\lambda_{D}/R$.

Note that from \eqref{eq:NLPB} and \eqref{eq:bc_CC} it follows that
\begin{equation}
\overline{\sinh {\phi }} = \dfrac{4\ell_{Du}}{ R}
\end{equation}

\begin{figure}[t]
\begin{center}
\includegraphics[width=0.99\columnwidth , trim=0.cm 0. 0.0cm
0.,clip=false]{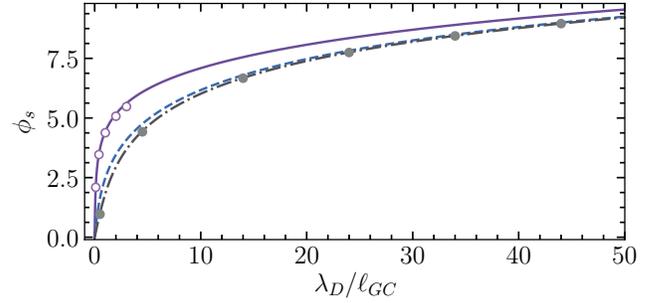}
\end{center}
\caption{Surface potential, $\phi_s$, as a function of $\lambda_{D}/\ell_{GC}$ calculated numerically using $R/\lambda_{D} \simeq 0.1,$ 1, and 10 (from top to bottom). Filled and open circles show calculations from Eqs.~\eqref{eq:grahame} and \eqref{eq:grahame_thin}. }
\label{fig:grahame}
\end{figure}

\subsection{Thick channel}

For a single wall or an infinitely large cylinder the relation between the surface potential and charge is given by the Grahame equation
\begin{equation}\label{eq:grahame}
  \phi_{s} = 2 \arsinh\left(\dfrac{\lambda_{D}}{\ell_{GC}}\right)
\end{equation}
In Fig.~\ref{fig:grahame} we plot $\phi_s$ as a function of $\lambda_{D}/\ell_{GC}$. It can be seen that Eq.\eqref{eq:grahame} remains very accurate for a thick channel and even when $\lambda_D/R =O(1)$ provided $\lambda_{D}/\ell_{GC}$ is large.

Substituting \eqref{eq:grahame} into \eqref{eq:cosh} one obtains $\overline{\cosh\phi} =  1$, which is equivalent to $\overline{\phi} =  0$ and $\overline{(\phi^{\prime} )^{2}} = 0$. However, this result holds only if $R \to \infty$. The data presented in Fig.~\ref{fig:cosh} show that $\overline{\cosh\phi}$ is well above unity at finite $\lambda_{D}/\ell_{GC}$, even for a thick channel, indicating that $\overline{\phi} \neq  0$.

The   calculation of $\overline{\phi}$  represents a challenge for a cylindrical geometry and we leave this for a future research. However, when the nanotube is weakly charged, $\phi_s \leq 1$, one can linearize Eq.~\eqref{eq:NLPB}:
\begin{equation}  \label{eq:DH}
\dfrac{1}{r} \dfrac{d}{d r} \left( r \dfrac{d \phi}{d r} \right) \simeq \lambda _{D}^{-2} \phi
\end{equation}
Integrating the above equation~\eqref{eq:DH} we obtain the potential profile:
\begin{equation} \label{eq:phi_DH}
\phi \simeq  \dfrac{2\lambda_{D}}{\ell_{GC}} \dfrac{I_0(r/\lambda_{D})}{I_{1}(R/\lambda_{D})},
\end{equation}
where $I_{0}$ and $I_{1}$ are the modified Bessel functions. The average potential is then
\begin{equation}\label{eq:average_phi_DH}
 \overline{\phi} \simeq \dfrac{4\ell _{Du}}{ R}
\end{equation}

It follows from \eqref{eq:grahame} that for weakly charged surfaces $\phi_s \simeq 2\lambda_{D}/\ell_{GC}$. Using \eqref{eq:zeta0} we then obtain
\begin{equation}\label{eq:average_phi_DH2}
\zeta_0 \simeq \dfrac{2\lambda_{D}}{\ell_{GC}}  \left(1 - \dfrac{2\lambda_{D}}{R} \right)
\end{equation}

\begin{figure}[t]
\begin{center}
\includegraphics[width=0.99\columnwidth , trim=0.cm 0. 0.0cm 0.,clip=false]{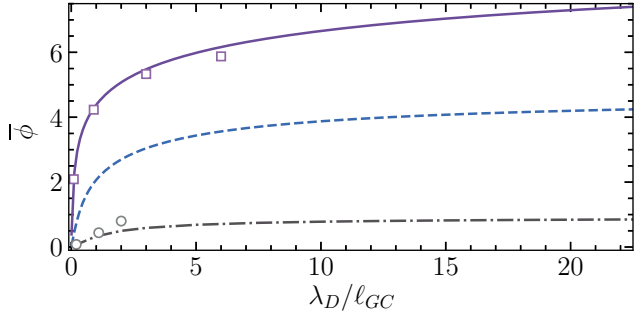}
\end{center}
\caption{$\overline{\phi}$ as a function of
 $R/\lambda_D$ computed for cylindrical channels of $R=10$ nm with fixed $\phi_{s} = 5$ and 1 (from top to bottom). Squares are obtained using Eq.~\eqref{eq:averyge_phi_thin}. Circles show predictions of linear theory Eq.~\eqref{eq:average_phi_DH}. }
\label{fig:phi}
\end{figure}

The numerical data presented in Fig.~\ref{fig:phi} confirm the validity of Eq.\eqref{eq:average_phi_DH} at $\lambda_D/\ell_{GC} \leq 1$ but show smaller $\overline{\phi}$ than this theoretical prediction when $\lambda_D/\ell_{GC}$ becomes larger. A calculation of the average potential and its effect on the zeta potential in the case of a highly charged thick nanotube remains very difficult and well beyond the scope of this paper. Nevertheless, we see that at large $\lambda_D/\ell_{GC}$ the surface potential (see Fig.~\ref{fig:grahame}) is much higher than $\overline{\phi}$, so that
the later can be neglected in the first-order calculations. Clearly, given this simplification
we do not expect the calculations of the zeta potential to be very accurate.

 It is naturally to assume that corrections to $\overline{\cosh\phi}$ and  $\overline{(\phi^{\prime} )^{2}}$ are proportional to the ratio of the EDL length to the channel area. Performing similar to described in \cite{vinogradova.oi:2021} calculations we obtain
\begin{equation}\label{eq:average_cosh}
\overline{\cosh\phi} \simeq  1 + \dfrac{4\lambda_D}{R}\left( \sqrt{1+\dfrac{\ell _{Du}}{\ell _{GC}}}-1\right),
\end{equation}
and
\begin{equation}\label{eq:average_dphi}
\overline{(\phi^{\prime} )^{2}} \simeq   \dfrac{8}{R \lambda_D} \left(\sqrt{1+\dfrac{\ell _{Du}}{\ell _{GC}}} - 1\right),
\end{equation}
Note that it follows then that these functions are related as
\begin{equation}\label{eq:average_cosh2}
\overline{\cosh\phi} \simeq  1 + \dfrac{\lambda_D^2 \overline{(\phi^{\prime} )^{2}}}{2}.
\end{equation}

Calculations from \eqref{eq:average_cosh} are shown in Fig.~\ref{fig:cosh}. It can be seen that Eq.\eqref{eq:average_cosh} gives a good match to the numerical data for thick nanotubes, and can even be used when $R \simeq \lambda_D$.

\subsection{Thin channel}

In the thin channel limit, $R/\lambda _{D}\ll 1$, the distribution of a potential in
the channel is approximately given by the parabola~\cite{silkina.ef:2019}
\begin{equation}  \label{eq:series}
\phi (r)=\phi _{s}+\dfrac{\sinh \phi _{s}}{4 \lambda _{D}^{2}}\left(
r^{2}-R^{2}\right).
\end{equation}
The relation between the surface potential and adsorbed charge reads
\cite{silkina.ef:2019}:
\begin{equation}  \label{eq:grahame_thin}
\phi_s = \mathrm{arsinh} \frac{4 \ell_{Du}}{ R}.
\end{equation}
The calculations from Eq.\eqref{eq:grahame_thin} are included in Fig.~\ref{fig:grahame}. It can be seen that \eqref{eq:grahame_thin} is accurate only when $\lambda _{D}/\ell_{GC}$ is small enough.

It follows then that the average potential is given by
\begin{equation}\label{eq:averyge_phi_thin}
\overline{\phi} \simeq \phi _{s} - \dfrac{R^2 \sinh \phi _{s}}{8 \lambda _{D}^{2}} \simeq \mathrm{arsinh} \frac{4 \ell_{Du}}{ R} - \dfrac{R}{2 \ell_{GC}}
\end{equation}

The calculation from Eq.\eqref{eq:averyge_phi_thin} are compared with numerical results in Fig.~\ref{fig:phi}. The fit is
quite good for $\lambda_D/\ell_{GC} \leq 3$, but at larger $\lambda_D/\ell_{GC}$ there is some discrepancy, and the average potential is higher than predicted by Eq.\eqref{eq:averyge_phi_thin}.

By direct integration of Eq.~\eqref{eq:series} using \eqref{eq:cosh} and \eqref{eq:grahame_thin} we can easily derive
\begin{equation}\label{eq:average_cosh_thin}
\overline{\cosh\phi} \simeq \sqrt{ 1 + \left( \dfrac{4\ell_{Du}}{R} \right)^2 }- \dfrac{2 \ell_{Du}}{\ell_{GC}}
\end{equation}
and
\begin{equation}\label{eq:series_dphi_thin}
\overline{(\phi^{\prime} )^{2}} \simeq  \dfrac{2}{ \ell_{GC}^{2}}
\end{equation}

\providecommand*{\mcitethebibliography}{\thebibliography}
\csname @ifundefined\endcsname{endmcitethebibliography}
{\let\endmcitethebibliography\endthebibliography}{}

\end{document}